\documentstyle[12pt,aaspp4]{article}

\begin{document}
\title{The Distribution of Barred Galaxies in the Virgo Cluster}
\author{Victor Andersen}
\affil{University of Alabama, Department of Physics and Astronomy}
\authoraddr{Box 870324, Tuscaloosa, AL, 35487-0324}

\begin{abstract}
A study of the distribution of barred and nonbarred disk galaxies in the 
Virgo cluster is presented in an attempt to use the frequency and spatial
distribution of galaxies with specific morphological 
features to study the efficiency of various environmental effects on the 
evolution of disk galaxies in clusters. The velocity distribution of the 
barred spirals
in the Virgo region is clearly different than that of the nonbarred spirals,
suggesting that barred spirals are more common in the main condensation 
of the cluster. A sample cleansed of galaxies not belonging to the main
cluster condensation using the subcluster assignments of
Binggeli et al.\markcite{members} (1993) bears this out, showing that the 
radial distribution
of barred spirals is more centrally condensed than that of nonbarred 
spirals. In contrast to the spiral galaxies, the distribution of barred 
S0 galaxies is statistically indistinguishable from that of nonbarred S0's.
Consideration of the level of tidal perturbation due to the cluster mass
distribution as compared to that due to individual galaxies suggests that
tidal triggering by the cluster mass distribution is the most likely 
source of the enhanced fraction of barred spirals in the cluster center.

\end{abstract}

\section{Introduction}

The galaxies in the cores of present day galaxy clusters are preferentially
found to be elliptical and lenticular galaxies, rather than spiral galaxies 
which predominant in lower density regions of the universe (Gisler 
\markcite{gisler} 1980;
Dressler \markcite{dress} 1980a). It seems unlikely that this is simply due 
to different galaxy
types forming in different environments, since observations of clusters at
redshifts of $z \sim 0.3-0.4$ show a much higher percentage of spirals 
in these 
clusters than in present day clusters (Couch et al. \markcite{couch} 1994; 
Dressler et al. \markcite{dress2} \markcite{dress3} 1994a-b). Apparently 
some environmental effect is responsible for
altering cluster spirals seen at high redshift beyond recognition as spirals
by the current day. Several possible mechanisms have been proposed; ram
pressure sweeping of the interstellar medium by the gas responsible for 
the cluster x-ray emission (Gunn \& Gott \markcite{gunngott} 1972), the 
cumulative effects of
galaxy-galaxy collisions in the dense cluster core (Richstone 
\markcite{rich} 1975), or
tidal effects due to the gravitational field of the cluster as a whole
(Merritt \markcite{merritt} 1983). Unfortunately to date, it has been 
difficult to distinguish
observationally between the various possible mechanisms.

\newpage

In a study of the Coma cluster, Thompson \markcite{comabars} (1981)
found that the percentage of barred galaxies within approximately 
$0.75 {\rm Mpc}$ of the cluster center ( $H_{0} = 75 {\rm km/sec/Mpc}$ and
a Virgo cluster distance of 20 Mpc is used throughout this paper )
was significantly higher than in the outer parts of the cluster. 
Thompson noted that this either meant that the excess of barred galaxies
represented a kinematically distinct component confined to the core of 
the cluster, or that bars were triggered by some mechanism as disk 
galaxies entered the cluster core. If the latter is true, the lifetime of
the induced bars must be the less than or the order of the core crossing
time; for the Coma cluster the core crossing time is approximately $10^{9} 
yr$, which is around 4--5 disk rotation times for a typical spiral galaxy.
Simulations of galaxy-galaxy interactions (Noguchi \markcite{noguchi2}
\markcite{noguchi} 1987, 1988) 
and interactions of a disk galaxy with 
a cluster gravitational field (Byrd \& Valtonen \markcite{byrdvalt} 1990) 
show that both
types of interaction can stimulate the formation of bars
in galaxies that would otherwise be stable against the 
development of a bar, and enhance the formation of bars in galaxies which
are already unstable to bar formation (Gerin, Combes \& Athanassoula 1990 
\markcite{gerin}). In his simulations, Noguchi\markcite{noguchi} (1988) 
finds that the lifetime of the induced bars is in the range 
$5 \times 10^{8}$--$1.5 \times 10^{9}yr$.  Thus, the frequency of barred 
galaxies in a given
cluster gives information about the interaction history of those 
galaxies in the recent past (i.e. over the last 3--6 disk revolutions
or so). The purpose of this project is to examine the distribution of barred
galaxies in the Virgo cluster, as an aid in discriminating 
between the relative importance of different environmental effects in 
cluster galaxies.

\section{The Virgo Cluster}

The proximity of the Virgo cluster gives it a unique advantage over other
clusters for the study of cluster galaxy morphology. As Dressler 
\markcite{morpcat} (1980b) has pointed out, in order to do morphological 
studies at the distance of most clusters requires high quality, high plate 
scale images taken with large reflectors. The relative nearness of the 
Virgo cluster means that the morphology of Virgo galaxies can be reliably
estimated using more easily accessible Schmidt plates.

In the Virgo cluster region, there exist several kinematically distinct
units, not all of which lie at the same distance as the cluster's primary
condensation (Tully \& Shaya 1984 \markcite{shaya}; Binggeli, Tammann \& 
Sandage 1987 \markcite{binggeli}; Binggeli, Popescu, \& Tammann 1993 
\markcite{members}). In this paper I will generally adopt the definitions 
and nomenclature for the various substructures given in 
Binggeli et al. (1993). 
The primary condensation of the Virgo cluster (the A cluster) is centered 
near the giant
elliptical galaxy M87, and contains the bulk of the elliptical and 
lenticular
galaxies in the cluster. The velocity histogram of the Virgo ellipticals
appears essentially Gaussian (cf. Figure 1), however M87 does not lie at
the peak of the velocity distribution for the central galaxies, and is
offset from the surface density maximum of the central galaxies as well.
Binggeli et al. (1987) 
argue that this means that the  core of the cluster is not in virial 
equilibrium, since if the mass distribution were centered on the surface
density maximum, M87 should be tidally truncated by the cluster core 
(Merritt\markcite{merritt}1983). The extended x-ray emission seen in the
cluster by ROSAT (B\"ohringer et al.\markcite{xray} 1994) is centered on M87,
suggesting that it does lie at the dynamical center of the cluster. 

The B cluster is a spiral dominated structure centered near the elliptical
galaxy M49. A determination of the distance to B puts it at the approximately 
the same distance as cluster A (Binggeli et al. 1993\markcite{members}). 
X-ray emission from cluster B has been detected using the ROSAT all sky 
survey data (B\"ohringer et al. 1994\markcite{xray}), suggesting that
even though the B cloud is at the distance of cluster A, it comprises a 
distinct kinematic unit from A. To the southwest of cluster B lie 
the W and W$^{\prime}$
clouds. The W cloud is at a higher velocity than cluster A, and distance
determinations indicate that it probably lies at about twice the distance
of A. The W$^{\prime}$ cloud lies between B and W on the sky, and also at a
distance and recession velocity intermediate between B and W. The M cloud 
lies to the 
west of  cluster A, and has a distance and redshift approximately twice that 
of A. Finally,  an extended structure lying to the south and at
around the distance of A  is the aptly named ``Southern Extension''.

\section{The Sample}

The initial sample consisted of all galaxies from the Virgo
cluster catalog (VCC) of Binggeli, Sandage \& Tammann \markcite{vcc} (1985) 
with total blue
apparent magnitudes $B_{T} \leq 14.0$. 
Galaxies which had recession 
velocities
$\geq 3000 {\rm km/sec}$ were excluded, since it is unlikely that
they are actual members of the Virgo cluster. 
Morphological types
were taken from the {\it Third Reference Catalogue of Bright Galaxies} (RC3)
(de Vaucouleurs et al. \markcite{rc3} (1991)), and the sample was divided
into elliptical, lenticular, and spiral $+$ irregular subsamples.
The elliptical subsample was retained for fiducial purposes only, since the 
ellipticals represent a population that define the core regions of the 
cluster. 
Galaxies in the disk galaxy (S0 and S$+$I) subsamples which were viewed to 
close to
edge on were excluded from the sample in order to avoid cases where 
distinguishing between barred and non-barred morphology is difficult. 
To determine at what inclination the determination of bar type became
difficult,
images of barred and non-barred galaxies at various
inclinations where examined. From this, it was determined that for galaxies 
with isophotal axis ratios $R_{25} \geq 2.5$ bar classification became
unreliable, so galaxies with axis ratios greater than 2.5 were
removed from the sample. This is also the approximate value
above which the fraction of galaxies classified as SAB and SB drops to
near zero, confirming that classification of bar types becomes difficult
for these more highly inclined galaxies.
 This separation left a final spiral sample of 
32 SA galaxies, 32 SB, and 26 SAB galaxies; and a lenticular sample with
32 S0 galaxies, 24 SB0 galaxies, and  4 SAB0 galaxies.

\section{Results}

If the Virgo cluster had little substructure,
analyzing the distribution of barred versus non-barred galaxies
could be done simply by looking at the fraction of different types of
galaxies at different radii from the cluster center. This was the approach 
used by Thompson \markcite{comabars} (1981) to detect the enhancement of
the barred galaxy fraction in the center of the Coma cluster. 
Given the complicated structure of the Virgo cluster, I chose initially to
examine the distribution of barred versus non-barred galaxies in 
velocity space, since most of the substructures are clustered to the high
end of the velocity distribution for galaxies in the VCC.

\begin{figure}
\plotone{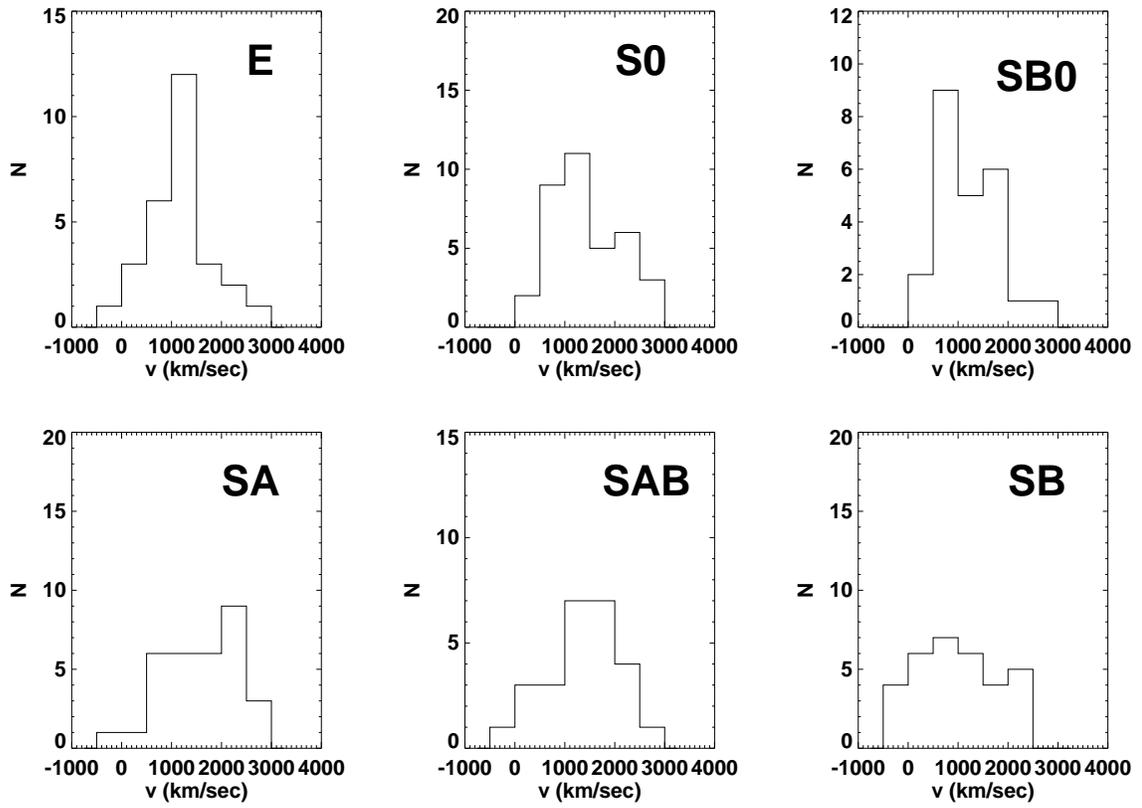}
\caption{Velocity Histograms for galaxy subsamples.}
\end{figure}

Velocity histograms for the various subsamples are plotted in Figure 1. The
histogram for the elliptical galaxies is plotted for comparison, since it
is characteristic of galaxies which are confined to the main condensation of
the cluster.
To formally test the difference between the velocity distribution of 
galaxies of 
different bar strength, the Kolmogorov-Smirnov (KS) test, which tests the 
hypothesis that two samples were drawn from the same underlying 
distribution, was employed. For the purposes of the KS test, the velocities
 of SB galaxies were compared to the combined velocities of the 
SA and SAB galaxies. This was done with the hope that the SB galaxies
would represent the population of most strongly and unambiguously barred
galaxies, and is also roughly consistent with the sample division used
by Thompson  for the Coma cluster. 
The KS test works by finding the 
maximum difference between the cumulative frequency distribution of the two 
samples, and then assessing the likelyhood that the difference is due to 
chance.  For the spiral galaxies, maximum deviation between 
the two samples is because of an excess of barred galaxies with
$v_{\sun} \leq 650$ km/sec, and the KS test finds that there is only
a 3\% probability that this difference is due to chance. Binggeli et al.
\markcite{members}(1993) have proposed that the velocity range 
$v_{\sun} \leq 500 {\rm km/sec}$ is the only range for the Virgo cluster
proper that is uncontaminated by 
interlopers, which suggests that the difference is due to an enhancement in 
the fraction of barred galaxies in the A cluster. 
In contrast to the spiral sample, the KS test finds a 60\% probability
that the S0 and SB0 galaxies are from the same population.

\begin{figure}
\epsscale{0.80}
\plotone{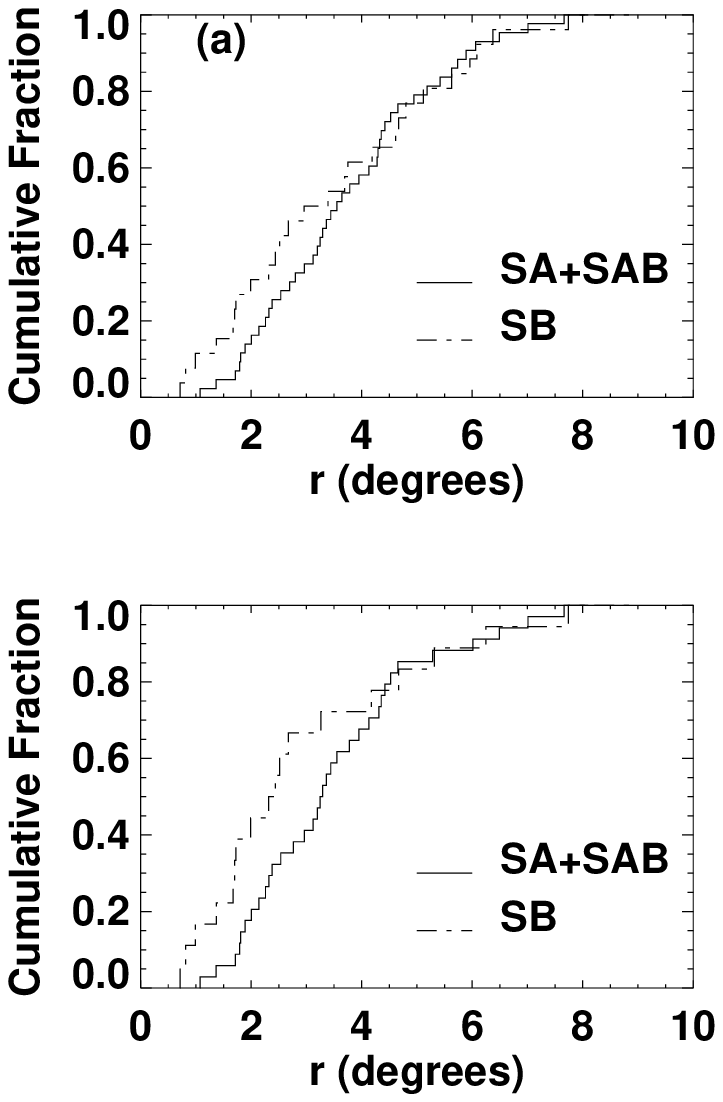}
\caption{Cumulative fraction as a function of radius for SA$+$SAB and
SB galaxies with (a) group B included and (b) excluded.}
\end{figure}

In order to study the spatial distribution of galaxies in the cluster, a 
method
is needed to identify galaxies which are not at the distance of the main
condensation and remove them from the sample. Ideally this would be done
using a distance indicator that does not depend on the galaxy's recession
velocity, such as the Tully-Fisher relation (Tully \& Fisher 
\markcite{tullyfisher} 1977) to determine the distance to each sample 
galaxy, and remove galaxies that were clearly at a different distance 
than the cluster. In practice, the lack of the appropriate data, or simply
the unsuitability of many sample galaxies (due to unfavorable inclination,
peculiarities in structure, etc. ) means that this approach is impossible
for the current sample. In addition, a purely distance dependent technique
would not identify kinematically distinct structures such as the B cluster
which lie at a common distance with the main cluster.
 Therefore, cluster membership was decided based
upon the assignments given in Binggeli et al. 
\markcite{members} (1993). The disadvantage of using this method is that
the criteria for assigning a galaxy to a particular substructure is somewhat
subjective.  Whenever possible for the current sample, 
the assignments 
have been verified using Tully-Fisher distances, and in every case have 
found to be accurate.

\begin{figure}[t]
\plotone{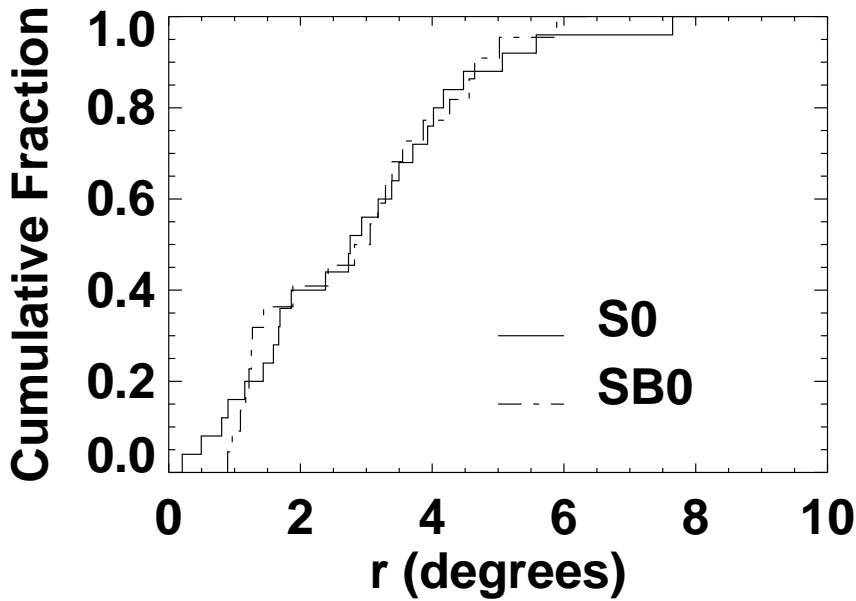}
\caption{Cumulative fraction as a function of radius for the S0 and SB0
galaxies. Only the case with group B included is plotted.}
\end{figure}
 
Using the Binggeli et al. \markcite{members} (1993) assignments, all 
galaxies from the background structures (i.e. groups M, W and W$^{\prime}$)
as well as the few in the southern extension were excluded from the cluster
sample. Whether to exclude galaxies assigned to the group
B is somewhat unclear. On the one hand it is apparently at the distance
of group A which suggests that it should be included as
belonging to the main cluster. On the other hand, both the redshift and
x-ray data indicate that B represents a kinematically distinct unit,
suggesting that the environment of the main cluster may not be the primary
external effect on component B's galaxies. Keeping this 
in mind, I analyzed the data for both the case of B included and 
excluded. 

The radial distribution of the cumulative fraction of galaxies with 
projected radius $\leq r$ for the spiral samples with and without group B
galaxies included is shown in Figure 2 (a-b). Both distributions show that 
the radial distribution of barred spirals is more centrally 
peaked than that of nonbarred spirals, with the discrepancy being more
pronounced for the case where group B galaxies are excluded. The KS test 
shows that in the case where B is included the two distributions would
be drawn from the same parent population 18\% of the time. For the 
case with B excluded, the probability of being drawn from the same 
population drops to 7\%. In either case, the discrepancy between the
two samples comes from higher fraction of barred galaxies within 
2--2.5$\arcdeg$ of the cluster center.

The radial distribution of cumulative fraction for the lenticular galaxies
in A and B combined is plotted in Figure 3. For the S0 galaxies the case 
with group B excluded was not examined separately, since the small number 
of lenticulars assigned to B (5) has a negligible effect on the radial
distributions. Unlike the case for the spirals, the radial distribution
of S0 and SB0 galaxies appear to be identical. The KS test agrees with
this qualitative impression, giving a 60\%
probability that the S0's andf SB0's are drawn from the same parent 
population.

\section{Discussion}

It seems likely that the increased fraction of barred spirals in the center
of both the Virgo and Coma clusters is due to these galaxies suffering 
strong tidal interactions in the inner parts of the cluster. Three immediate
possibilities present themselves for how the bars are formed; (1) Tidal
triggering due to the gravitational field of the cluster, (2) tidal 
triggering due to encounters between individual galaxies in the cluster,
and (3) Thompson (1980)\markcite{comabars} suggested that tidal stripping of
the galaxies' dark matter halo would lead to a 
puffing up of the remaining halo; the reduced central mass density in the
galaxy may then leave the galaxy disk unstable to spontaneous bar formation
(Ostriker \& Peebles 1973\markcite{ostriker}). However Byrd \& Valtonen 
(1990)\markcite{byrdvalt}
find that a bar is induced in a galaxy long before significant stripping 
occurs. For this reason I will not consider the third mechanism in the
following discussion.
The question then becomes whether the tidal effects
are due to the concentrated mass of the cluster itself, or due to an
enhancement of the galaxy--galaxy interaction rate because of the increased
galaxy density in the cluster core. It is not possible to use detailed
considerations to address this subject at present. Although simulations
exist that demonstrate both types of interactions are capable of inducing
bars in some circumstances, many parameters play a role in 
determining whether a bar is formed or not (perturbation strength, 
ratio of halo (or bulge) to disk mass,
sense of encounter with respect to disk rotation axis (prograde or 
retrograde)), and the extensive 
grids of models that would delineate the regions
of parameter space where interactions do and don't lead to bar formation
do not exist. The best it is possible to do at present is to use the 
existing models to point out gross results as to what parameter values
seem to control bar formation in a given interaction.

In their simulations of the formation of ocular
spirals, Elmegreen et al.\markcite{ocular} (1991) characterize the strength
of the perturbation using the parameter $S$ which is ratio of the change
of momentum due to the perturbation to the initial momentum for a particle
at the outer edge of the galaxy.
Elmegreen et al. found bars were formed in their simulations only if the
value of $S$ exceeded a threshold value, showing that for a given galaxy,
the strength of the perturbation is a controlling factor in whether a bar
is formed or not. 

The utility of $S$ in estimating perturbation strengths
for observed galaxies is not great however, since it depends on the ratio
of the timescale over which the perturbation acts to the orbital timescale
of a star in the outer part of the galaxy's disk, making $S$ difficult
to derive from observations. For this reason, it is common to use a 
parameter which isolates those quantities which are more easily derived
from observations, such as masses and distances. Byrd \& Valtonen 
(1990)\markcite{byrdvalt} define the parameter $P$, which for an arbitrary
mass distribution on the part of the perturber depends on the local gradient
of the gravitational force due to the perturber $\nabla f$ as $P=\nabla f
r_{g}^{3}/(G M_{g})$, where $r_{g}$ is the optical radius of the galaxy, and
$M_{g}$ is the mass of the galaxy within the optical radius.
For a point mass
perturber this simplifies to $P = (M_{p}/M_{g})(r_{g}/r_{p})^{3}$.
As in the simulations of Elmegreen et al., bars are formed in 
Byrd \& Valtonens simulations only when the perturbation level (measured 
using $P$ in this case) exceeds some threshold level, generally in the 
range 0.006--0.1, where the lower value is for a disk with no massive halo,
and the upper value for the case where the halo mass dominates the total
galaxy mass ($M_{halo}/M_{disk} \gtrsim 2$).

The self gravity of a galaxy's disk is
important in determining whether a bar is formed or not (Noguchi 
\markcite{noguchi2} 1987). This means that
the lower the ratio of disk to halo (or bulge) mass, the more stable a 
galaxy will be against bar formation (Ostriker \& Peebles 
\markcite{ostriker} 1973). Mihos \& Hernquist (1994)\markcite{mihos} have
simulated interactions using galaxies with dense bulges and galaxies
with no bulge component. They find that the presence of a sufficiently dense
bulge does indeed inhibit the formation of a bar in cases where a strong
bar is formed in the case of a pure disk galaxy.

Encounters that occur with too high
a velocity should have little effect on the structure of galaxy disks, since 
the impulsive force on disks stars is small because of the short duration
of the interaction. Retrograde encounters are less effective than prograde
encounters in affecting the structure of disk galaxies for the same reason;
in the rest frame of disk stars, the combination of disk rotation velocity 
plus encounter velocity means that a star sees the tidal impulse for only
a relatively short time. For a given galaxy then, a minimum level of 
perturbation is necessary to induce bar formation, modified by the details
of the galaxy and encounter. In this way, a certain level of perturbation
is a necessary but not sufficient condition for a bar to be induced.

In an attempt to determine whether barred spirals in Virgo where more 
likely to be interacting with other galaxies more often than nonbarred 
cluster spirals, the VCC was searched down to 
a limiting magnitude of $B_{T}=18.0$, the nominal completeness limit of 
the catalog, to identify the nearest neighbor for each spiral in the 
main cluster.  Galaxies which were identified as non-cluster members in 
Binggeli et al. \markcite{members} (1993) were rejected from consideration.
Galaxies near the center tend to have closer nearest neighbors than galaxies
at greater radius, simply due to the increase in galaxy surface density near
the cluster center. In order to remove this effect the local surface density
near each galaxy was evaluated using the surface density fit to the VCC data
given in Ferguson \& Sandage (1990)\markcite{ferguson}. The local surface 
density $\rho$ was then used to calculate the expected distance to the 
nearest neighbor assuming a Poisson random distribution distribution of 
galaxies 
$d_{Poiss} \approx (2\sqrt{\rho})^{-1}$ (see e.g. Keel \& van Soest 1992)
\markcite{keel}. Although the assumption of a random distribution is not 
strictly correct since the surface density varies with radius, the fact that
core radius of the fit distribution of $1.4\arcdeg$ is much larger than even
the greatest nearest neighbor separation means that on the scales being
considered, the surface density is close to constant so that the assumption
used to find the expected nearest neighbor distance is valid.

\begin{figure}[t]
\plotone{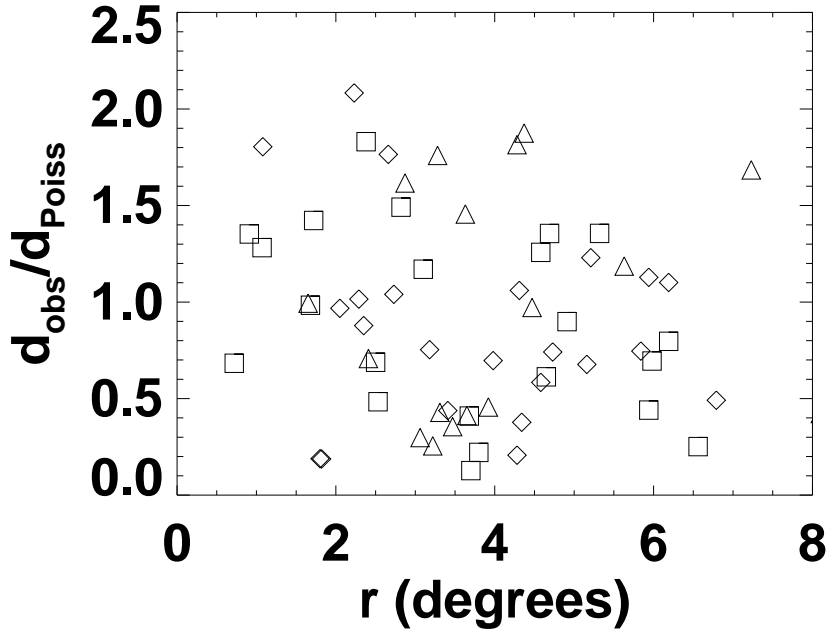}
\caption{Ratio of observed nearest neighbor distance to expected distance
given the local surface density, plotted against cluster--centric radius. SB 
spirals are plotted as squares, SAB as triangles, and SA as diamonds.}
\end{figure}

The ratio of the measured nearest neighbor distance to $d_{Poiss}$
is plotted against the radius in Figure 4. Two results are noteworthy from
this graph; first, as an ensemble the SB spirals are no more likely to
have a nearer neighbor than SA and SAB spirals, and second, there is no
tendency for galaxies near the center of the cluster, the region of 
increased bar fraction, to have nearby
companions. It is important to note that although the nearest neighbor 
analysis shows that the barred cluster spirals do not preferentially have
nearby companions, it does not demonstrate unambiguously that galaxy--galaxy
interactions is not a viable mechanism for stimulating bar formation. 
The high density and relative velocities of the galaxies in the clusters
core means that in an unbound encounter, the time spent within $d_{Poiss}$ 
by the perturbing galaxy is much shorter than the lifetime of the bar. 

What really is needed then is a measure of the efficiency of galaxy 
collisions in the inner parts of the cluster for stimulating bars. In
specific, a measure of how likely a bar inducing collision is for a
single crossing of the cluster core. If the expected time between bar
inducing collisions is of the order of the core crossing time then collisions
may be able to compete effectively with cluster tides in the formation of
bars, if the collision time is much longer than the crossing time then 
cluster tides should dominate. For a region of linear dimension $d$ 
with a galaxy density $n$, each with a cross section $\sigma$ for bar 
forming collisions, a galaxy traveling at a velocity $v$ will have collision
and crossing times of $t_{coll}=1/n\sigma v$ and $t_{cross} = d/v$, so the 
ratio of the two times is just \begin{equation} t_{cross}t_{coll}^{-1} =
n \sigma d. \end{equation}

If an induced bar has a lifetime that is much longer than the orbital time
scale of the galaxy in the cluster, the chance of forming a bar depends not
only on $t_{cross}t_{coll}^{-1}$, but on the number of times the galaxy has 
crossed the cluster core. That is, if the galaxy has crossed the core $N$ 
times, galaxy-galaxy collisions will be efficient at forming bars if
\begin{equation} t_{cross}t_{coll}^{-1} \gtrsim N. \end{equation} The fact 
that $\sim 2/3$ of spiral galaxies are of type SB or SAB may imply that in
general bars in galaxies are long lived phenomena. On the other hand, there
are indications that collisionally induced bars may not always be long lived.
In his simulations of bar inducing interactions, Noguchi\markcite{noguchi2}
(1988) finds that the bars have lifetimes on the order of 4--6 disk rotations.
The dissolution of the bar in these simulations coincides with the time it 
takes for gas clouds comprising a few percent of the galaxies initial disk
mass 
to make it into the inner parts of the galaxy. Simulations by Norman, 
Sellwood, \& Hasan\markcite{norman} (1995) to explore the effect of 
increasing the central mass concentration in models which contain a strong, 
self-consistent bar show that accumulation of $\sim 5$\% of the initial 
disk mass in the galaxy core leads to rapid dissolution of the bar. The
remnants of the bar end up in a distribution similar to a galactic bulge.
Although they do not follow the galaxies in their simulations to bar 
dissolution, Byrd \& Valtonen\markcite{byrdvalt}(1990) see large mass
inflows into the central $1 \rm kpc$ of their galaxies, suggesting that
the lifetime of the bars in their simulations may also be limited. For this
reason the following discussion will assume that whatever mechanism triggers
the bar formation in the cluster galaxies must operate efficiently over a
single core crossing. 

If for the purposes of this argument we consider
a galaxy on a radial orbit through the cluster, then the average density
within a radius $d/2$ of the center of the cluster can be obtained using
the density profile from Ferguson \& Sandage (1990)\markcite{ferguson}.
Doing this, the ratio of time scales becomes \begin{equation} t_{cross}
t_{coll}^{-1} = 6 n_{\circ} \sigma r_{c} f(d/2r_{c}), \end{equation}
where $n_{\circ} = 175 {\rm galaxies/Mpc^{3}}$ is the central density of
galaxies brighter than $B_{T}=14.0$, $r_{c}=489 {\rm kpc}$ is the core 
radius of the density distribution (for a Virgo distance of 20 Mpc),
and the function $f$ is $f(x)=(\ln(x+(x^{2}+1)^{1/2})-x/
(x^{2}+1)^{1/2})/x^{2}$. A radial orbit was selected not only for 
computational simplicity, but also because it will maximize the
value of $t_{cross}t_{coll}^{-1}$ since it probes the highest density 
regions of the cluster. 

The condition for collisional excitation of bars to be an effective process
is then for the cross section for bar formation to be high enough that
$t_{cross}t_{coll}^{-1} \gtrsim 1$. The cross section may be related to the 
perturbation strength $P$ by assuming that the cross section is related to
the impact parameter $b$ as $\sigma = \pi b^{2}$. If we assume that the 
colliding galaxies are of equal mass, then $P=(r/b)^{3}$ where $r$ is the
optical radius of the galaxy. In their study of cluster tidal effects, 
Byrd \& Valtonen (1990)\markcite{byrdvalt} find that triggering occurs at 
perturbation levels in the range $0.006 \leq P \leq 0.1$, where the lowest 
value is associated with systems with little or no dark matter halo, while 
the higher value is for systems with halo masses greater than about two 
times the disk mass. Figure 5 shows the ratio of time scales
plotted against radius for three different values of $P$, where the cross 
section has been derived for a galaxy radius of 10 kpc. The vertical lines
are the radius at which the perturbation strength due to the cluster mass is
the same as that for the curve of $t_{cross}t_{coll}^{-1}$ plotted with the
same line style.  The dashed line is for $P=0.1$ showing that if
cluster galaxies have a significant dark halo, galaxy-galaxy collisions
should not be an efficient bar stimulating mechanism. The solid lines
are for a value of $P=0.043$, which was chosen because it was the 
perturbation level due to the cluster at the radius  within which
there is an enhanced bar fraction, in this case within $2\arcdeg=700 
{\rm kpc}$. Again in this case, $t_{cross}t_{coll}^{-1}$ is still 
significantly less than 1. Finally, the dashed--dot line is plotted for
a value of $P=0.005$, which is comparable to Byrd \& Valtonen's lowest
triggering level, and was selected since it gives $t_{cross}t_{coll}^{-1} 
\approx 1$
at a core radius. Although this condition does raise $t_{cross}t_{coll}^{-1}$
near the center, it also makes the region over which the cluster reaches the
same triggering level over 2 Mpc in radius, much larger than the region
of enhancement that is actually observed.
Although the above arguments do not establish unambiguously that cluster
tidal interactions are the dominant mechanism, they are at least strongly
suggestive that this is the case. 

\begin{figure}[t]
\plotone{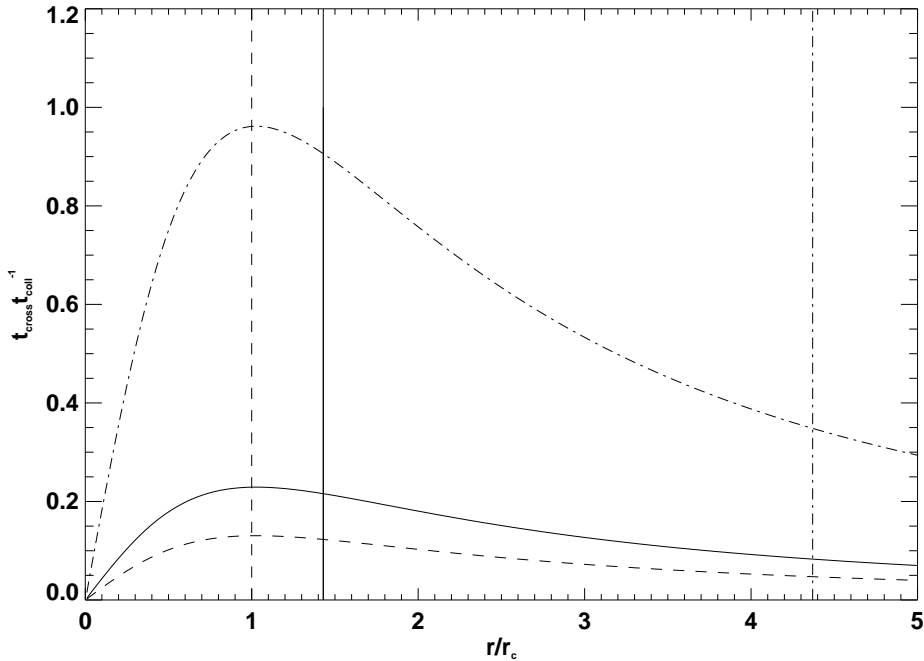}
\caption{$t_{cross}t_{coll}^{-1}$ vs. radius for different perturbation
levels. The vertical lines give the radius within which the cluster mass
gives the level of perturbation for which the curves were calculated.}
\end{figure}

Whether cluster tides or galaxy-galaxy interactions is the responsible
mechanism for inducing bars, the fact that a triggered bar may only persist 
for a few disk 
rotations suggests the interesting possibility that a single galaxy may 
undergo repeated episodes of bar formation and dissolution each time it
orbits through the cluster core. Since the bar remnants form a structure
similar to a galaxy bulge (Norman et al. 1995)\markcite{norman}, this 
may be a mechanism by which the morphology--density relation 
(Gisler  \markcite{gisler} 1980; Dressler \markcite{dress} 1980a) arises.
While this is clearly speculative at this point, the possibility definitely
warrants further investigation. 

\subsection{Comparison with the Coma Cluster}

The increased fraction of barred spirals in the core of the Virgo cluster
is in qualitative accordance with Thompson's finding for the Coma cluster,
however a more detailed comparison between the results for the two clusters
is instructive. The cluster--centric radius over which
there is an enhancement in bar fraction is essentially the same for both 
clusters, around
$0.75 {\rm Mpc}$. This is somewhat puzzling if the cluster mass is inducing 
the bars, since Coma is more massive than Virgo, so that a given level of 
perturbation should occur at a greater radius in Coma than in Virgo. 
Using the mass determination for Coma given in Hughes \markcite{comamass}
(1989), there is $2.67$ times as much enclosed mass within $0.75 {\rm Mpc}$ 
as in Virgo. This means that a galaxy in Coma will feel a level of 
perturbation corresponding to that felt in Virgo at a radius of 
$0.75 {\rm Mpc}$ at a radius of $\sim 1.2 {\rm Mpc}$. Given the limited 
sample sizes, the uncertainties in the extent over which there is an 
enhancement in bar fraction are probably large
enough that this discrepancy is not highly significant.
Careful study of several more nearby clusters will be
necessary to test whether this difference is meaningful.

Unlike in Coma, where the fraction of SB0 galaxies is also higher in the 
cluster
center, Virgo shows no such enhancement. This difference could potentially 
be due to the relatively higher mass of the Coma cluster. Since a massive
central spheroidal component in a galaxy can help stabilize a galaxy against
bar formation, a larger perturbation is required than for a galaxy
with a smaller bulge. Given that S0's have more massive bulges than 
intermediate and late type spirals (Simien \& de Vaucouleurs 1986
\markcite{simien}), it should be more difficult to trigger bar formation
in S0 disks than in the disks of spirals.
Perhaps we are seeing a case where the mass 
concentration in Virgo is large enough to trigger bars in spirals but not
S0 galaxies, while Coma's greater mass may be able to trigger bars in both
spirals and S0's. This result is also speculative with the current
data, analysis of more clusters may also be helpful in examining this
point more thoroughly.  

\section{Conclusions}

A study of the velocity distribution of disk galaxies in the Virgo cluster
region has been carried out, with the result that velocity distribution of
barred spirals is skewed to lower recession velocities when compared to 
weakly or unbarred spirals. Since galaxies with low velocities in the Virgo
region are preferentially associated with the main condensation of the 
Virgo cluster, this suggests the difference in the velocity distribution
for spirals of different bar strength is due to an increase in the fraction
of barred spirals associated with the main cluster when compared to less
dense structures in the region. 

Isolation of galaxies which belong to Virgo's main condensation indicate
that the radial distribution of SB galaxies is more centrally peaked than 
that of the SA and SAB galaxies. 
Furthermore, there is no indication that barred spirals have more nearby
companions than do unbarred spirals. Although the lack of nearby companions
by itself does not exclude the possibility that galaxy-galaxy interactions
are the dominant bar triggering mechanism, the fact that the ratio of
crossing to collision times is much less than unity for reasonable
perturbation strengths suggests that galaxy collisions are not efficient
enough to be responsible for the increased bar fraction.
The inference from these results is
that the central concentration of barred galaxies is due to triggering of
bars by the cluster mass distribution, as suggested by the simulations
of Byrd \& Valtonen (1990)\markcite{byrdvalt}.

Unlike the spiral galaxies, the S0 and SB0 galaxies in Virgo have velocity
and spatial distributions that are indistinguishable from one another. 
It is speculated that the difference between the spiral and lenticular
galaxies may be due to the cluster tidal forces being insufficient to over
come the stabilizing effect of the S0's more massive central bulges. 
Taken together, the enhanced fraction of barred galaxies in the center of
the Virgo and Coma clusters is compelling evidence for the importance of 
tidal interactions in the evolution of cluster galaxies.

\acknowledgments A careful reading of an earlier draft of this paper by
Bill Keel resulted in significant improvement in its presentation.

\end{document}